\documentstyle[prl,aps,epsf]{revtex}
\input epsf
\begin{document}
\draft
\twocolumn[
\hsize\textwidth\columnwidth\hsize\csname @twocolumnfalse\endcsname

\title{ Stable vortex-antivortex molecules in mesoscopic superconducting
triangles }
\author{V.~R.~Misko$^{a,}$\cite{A1}, V.~M.~Fomin$^{a,}$\cite{A2},
J.~T.~Devreese$^{a,}$\cite{A3}, V.~V.~Moshchalkov$^{b}$}
\address{$^{a}$Theoretische Fysica van de Vaste Stoffen, 
Universiteit Antwerpen (U.I.A.), \\
Universiteitsplein 1, B-2610 Antwerpen, Belgi\"e \\
$^{b}$Laboratorium voor Vaste-Stoffysica en Magnetisme, 
Katholieke Universiteit Leuven, \\
Celestijnenlaan 200 D, B-3001 Leuven, Belgi\"e}
\date{\today}
\maketitle

\begin{abstract}
A thermodynamically stable vortex-antivortex pattern has been
revealed in mesoscopic type I superconducting triangles, contrary
to type II superconductors where similar patterns are unstable.
The stable vortex-antivortex ``molecule'' appears due to the
interplay between two factors: a repulsive vortex-antivortex
interaction in type I superconductors and the vortex confinement
in the triangle.

\end{abstract}

]

Symmetrically-confined vortex matter in superconductors,
superfluids and Bose-Einstein condensates offers unique
possibilities to study the interplay between the $C_{\infty }$
symmetry of the magnetic field and the discrete symmetry of the
boundary conditions. More specifically, superconductivity in
mesoscopic equilateral triangles, squares etc. in the presence of
a magnetic field nucleates by conserving the imposed symmetry
($C_{3}$, $C_{4}$) of the boundary conditions\cite{chib} and the
applied vorticity. As a result, in an equilateral triangle, for
example, in an applied magnetic field $H$ generating two flux quanta, $%
2\Phi _{0}$, superconductivity appears as the $C_{3}$-symmetric combination $%
3\Phi _{0}-\Phi _{0}$ (further on denoted as ``$3-1$'') of three vortices
and one {\it antivortex} in the center. These symmetry-induced antivortices
can be important not only for superconductors but also for symmetrically
confined superfluids and Bose-Einstein condensates. Since the order
parameter patterns reported in Refs.~\onlinecite{chib} have been obtained in
the framework of the linearized Ginzburg-Landau (GL) theory, this approach
is valid only close to the nucleation line $T_{c}(H)$. Can then these novel
symmetry-induced vortex-antivortex patterns survive deep in the
superconducting state? Several attempts have been already made to answer
this crucial question. In the limit of an extreme type II superconductor ($%
\kappa \gg 1$), it has been shown that for a thin-film square, a
configuration of one antivortex in the center and four vortices on the
diagonals of the square is unstable away from the phase boundary \cite{bk}.
According to the analysis based on the coupled nonlinear GL equations, the
vortex-antivortex pairs are unstable and no antivortices appear
spontaneously at the $T,\ H$ points far away from the $T_{c}(H)$ line\cite%
{bp}. Possible scenarios of penetration of a vortex into a mesoscopic
superconducting triangle with increasing magnetic field have been studied in
Ref.~\onlinecite{tricrete}. Two different states were considered: a single
vortex state and a state in the form of a symmetric combination of three
vortices and an antivortex with vorticity $L_{av}=-2$ (``$3-2$''
combination). The calculations\cite{tricrete} have shown that while a single
vortex enters the triangle through a midpoint of one side, the ``$3-2$''
combination turns out to be energetically favorable when the vortices are
close to the center of the triangle. Equilibrium is achieved when a single
vortex is in the center of the triangle. When approaching the phase
boundary, the free energy of a single-vortex state tends to coincide with
the free energy of the ``$3-2$'' combination\cite{tricrete}, thus confirming
conclusions \cite{chib,bk} that formation of antivortices is possible in the
close vicinity to the phase boundary.

The previous inferences \cite{bk,bp} on vortex-antivortex states in
mesoscopic structures seem to give us no hope to find {\it stable}
vortex-antivortex configurations deeper in the superconducting state,
considering them just as the features appearing in materials with $\kappa
\gg 1$ at the phase boundary together with superconductivity. {\it Here we
propose the new solution demonstrating the stability of the
vortex-antivortex patterns.} This solution is based on the simple conjecture
made by one of the authors (VVM, \cite{vvmunp}): the main source of the
vortex-antivortex pattern instability, namely vortex-antivortex {\it %
attraction}, can be removed by taking -- instead of type II --
type I superconductors, where vortex-antivortex interaction
becomes repulsive. Indeed, when passing through the dual point
$\kappa=1/\sqrt{2}$, the vortex-vortex interaction changes the
sign\cite{brandt,brass,blatter} and becomes attractive at
$\kappa<1/\sqrt{2}$. At the same time, the vortex-antivortex
interaction becomes repulsive. Therefore, one can expect that
presence of {\it antivortices}, together with {\it confinement} of
vortices and antivortices due to a potential barrier at the
boundaries, can stabilize novel vortex-antivortex patterns in a
mesoscopic sample of type I superconductor. Optimizing the
geometry and the sizes of mesoscopic samples, one can therefore
fulfil the conditions necessary for the existence of stable
vortex-antivortex configurations. For instance, presence of sharp
corners is known\cite{hml,vg,wedel,monat,wesql} to lead to a
strongly inhomogeneous distribution of the superconducting order
parameter in a mesoscopic sample. Enhanced superconducting
condensate density at the corners prevents vortices from leaving
the sample. Altogether, a triangular type I superconducting sample
seems to be an appropriate candidate to search for a stable
vortex-antivortex configuration.

\begin{figure}
\protect\centerline{\epsfbox{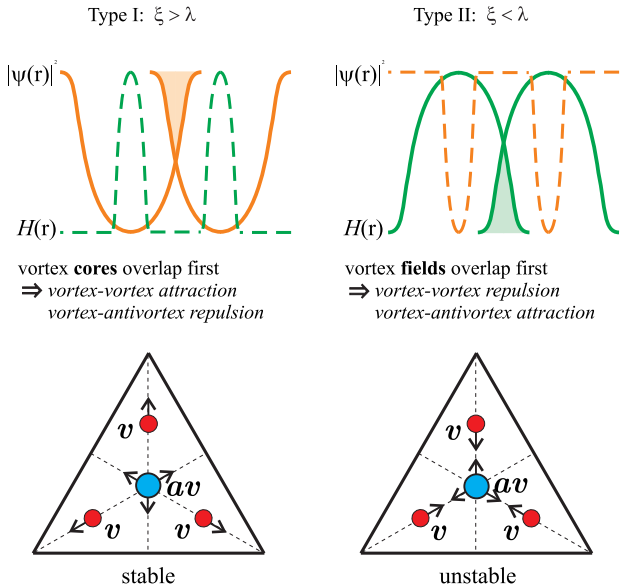}}
\smallskip 
\caption{
Schematics of unstable (type~II) and stable (type~I)
vortex-antivortex molecules in a mesoscopic superconducting triangle.
}
\label{Fig.1}
\end{figure}

Fig.~1 further illustrates this idea. In a type I superconductor with $\xi
>\lambda$, vortex cores overlap first when vortices approach each other.
This triggers vortex-vortex attraction (and vortex-antivortex repulsion). In
contrast with that, in a type II superconductor with $\xi <\lambda$, local
fields created by vortices overlap first when vortices approach each other,
thus inducing vortex-vortex repulsion (and vortex-antivortex attraction). As
a result, a vortex-antivortex combination, consisting of three vortices ($%
L_{3v}=3$) and one antivortex with vorticity $L_{av}=-1$ (``$3-1$''
combination, or $3v+1av$ {\it molecule}) is stable or unstable,
respectively, in a mesoscopic type I or type II superconducting triangle.

To verify these intuitive considerations we investigate a
mesoscopic equilateral triangle of a type I superconductor in an
applied magnetic field. The side of the triangle is chosen
($a=1\mu$m) to be larger than both characteristic lengths of the
GL theory, $\xi$ and $\lambda$. In order to provide a sample to be
a type I superconductor, we should consider in fact a triangular
{\it prism} with a height $h \gg \xi$ (and also $h \gg \lambda$).
This allows us to avoid an increase of $\kappa$ resulting in a
known effect when a mesoscopic sample made of material, which is
type I
superconductor in bulk, becomes effectively type II superconductor \cite%
{tricrete,monat,wesql,tinkhis}. For comparison, a mesoscopic triangle of
type II superconductor will be also considered. In our calculations, we use
the GL parameters of Pb (type I): $\xi_{\mbox{\footnotesize Pb}}=82$~nm, $%
\lambda_{\mbox{\footnotesize Pb}}=39$~nm, $\kappa_{\mbox{\footnotesize Pb}%
}=0.48$, and of Nb (type II): $\xi_{\mbox{\footnotesize Nb}}=39$~nm, $%
\lambda_{\mbox{\footnotesize Nb}}=50$~nm, $\kappa_{\mbox{\footnotesize Nb}%
}=1.28$ \cite{poole}.

In the description of the superconducting properties of mesoscopic triangles
we rely upon the GL equations for the order parameter $\psi$ and the vector
potential {\bf A} of the magnetic field $\mbox{\bf H}=\mbox{rot}\mbox{\bf A}$%
\cite{tinkhis,poole,gl50}. In the dimensionless form, when keeping the
temperature dependence explicitly, the GL equations are
\begin{eqnarray}
\left(-i\nabla-\mbox{\bf A}\right)^{2}\psi- \psi\left[\left(1-\frac{T}{T_{c}}%
\right)-\mid\psi\mid^{2}\right]=0,  \label{gledpsi} \\
\kappa ^{2} \Delta\mbox{\bf A}=\frac{i}{2}\left(\psi^{\ast}\nabla
\psi-\psi\nabla\psi^{\ast}\right)+\mbox{\bf A}\mid\psi\mid^{2}.
\label{gleda}
\end{eqnarray}
Here the GL parameter $\kappa=\lambda(T)/\xi(T)$, and $\xi(0)$ serves as the
unit length. The imposed boundary condition is
\begin{equation}  \label{bcd}
\mbox{\bf n}\cdot \left. \left(-i\nabla-\mbox{\bf A}\right)\psi \right|_
{boundary}=0.
\end{equation}
Topological characteristics of solutions of the GL equations are
determined by (anti)vortex core lines. One revolution along any
closed path around such a line changes the phase of the order
parameter by $2\pi L$, where $L$ is the winding number (vorticity)
of a vortex or antivortex.

Close to the sample, the local magnetic field is distorted as compared to
the applied magnetic field $\mbox{\bf H}$. However, far away from the sample
the distortion is negligible, and the symmetric gauge of the vector
potential can be applied: 
$\mbox{\bf A}=\left[\mbox{\bf H}\times\mbox{\bf r}\right]/2.$
The boundary condition, which equates the vector potential of the local
field with that of the applied field, is justified at the boundaries of the
simulation region, which is taken large enough to provide that all the
changes of the magnetic field occur {\it inside} this region (see, for
example, Refs.~\onlinecite{tricrete,dspg,webr}). Further on, we choose the $z
$-axis coinciding with the direction of the applied magnetic field, which is
normal to the triangular base. Because $h \gg \lambda$, a non-zero $z$%
-component of the magnetic field is appreciable in the vicinity of the
bases, where the field skirts the prism partially penetrating it. Near the
central cross-section in the $xy$-plane, the magnetic field and the order
parameter are uniform in the $z$-direction, and the problem becomes
effectively two-dimensional. The third dimension is taken into consideration
by imposing that, inside the simluation region, the total magnetic flux
through any cross-section, perpendicular to the $z$-axis, is the same\cite%
{dspg,webr,akker}. The GL equations (\ref{gledpsi}), (\ref{gleda}) with the
above boundary conditions are solved numerically, using the
finite-difference method, on a square mesh with the density of 200 nodes per
side of the triangle.

The calculations are performed for temperature $T/T_{c}=0.96$.
First, we find the magnetic field regions, in which states with
antivortices are expected to take place in a type I
superconducting triangle. The free energy calculations show that
at low fields, from $H_{0}=0$ to $0.14H_{c}(0)$, where $H_{c}(0)$
is the thermodynamical critical field\cite{tinkhis,poole} at zero
temperature, no vortices penetrate the triangle. The order
parameter distribution changes continuously from homogeneous at
$H_{0}=0$ to a strongly inhomogeneous one, characterized by
concentration of the superconducting phase in the corners. At
$H_{0}\approx0.14H_{c}(0)$ one vortex enters the triangle. For
vorticity $L=1$, the equilibrium is achieved when the vortex is in
the center of the triangle. For magnetic fields within the region
from $H_{0}=0.25$ to $0.38H_{c}(0)$, a giant vortex state with the
total vorticity $L=2$ is energetically preferable. Further on, we focus on $%
H_{0}=0.32H_{c}(0)$, which corresponds to the states with the total
vorticity $L=2$. This state can be represented by two possible
configurations: (i) two vortices in the form of a multivortex or a giant
vortex state; (ii) a symmetric combination of three vortices on the triangle
bisectors and one antivortex in the center (``$3-1$'' combination, or $3v+1av
$ molecule). (Symmetric combinations with a larger number of vortices and
antivortices such as ``$6-4$'', ``$9-7$'' etc. possess a higher free energy
than the ``$3-1$'' combination and are not considered.)

\begin{figure}
\protect\centerline{\epsfbox{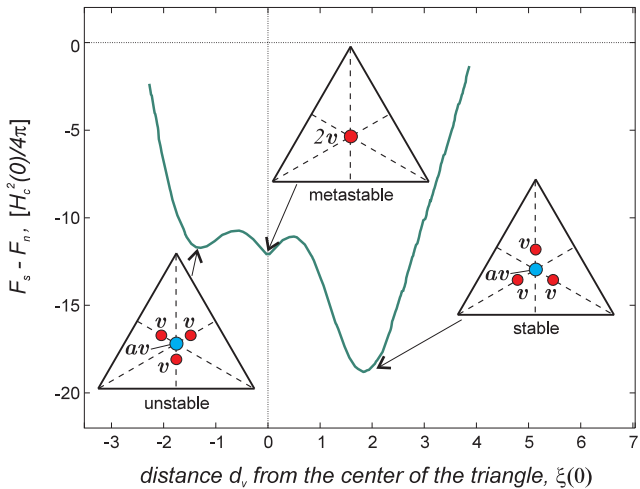}}
\smallskip 
\caption{
The free energy $F_{s}-F_{n}$ [measured in $%
H_{c}^{2}(0)/4\pi$] as a function of the distance $d_{v}$ from the center of
a mesoscopic type I superconducting triangle for a symmetric combination of
three vortices and one antivortex (the ``$3-1$'' combination, or $3v+1av$
molecule), at $T/T_{c}=0.96$, $H_{0}=0.32H_{c}(0)$, $\kappa=0.48$.
}
\label{Fig.2}
\end{figure}

According to our calculations, it is the ``$3-1$'' combination that
minimizes the free energy in case of a type I superconductor. In Fig.~2, the
free energy for this combination is shown as a function of the distance $%
d_{v}$ counted from the center of the triangle along the bisectors to
vortices. There are three minima of the free energy as a function of $d_{v}$%
. The first minimum, which is at $-1.29\xi(0)$ from the center of the
triangle, corresponds to a configuration when vortices are situated between
the center of the triangle and the midpoints of the sides of the triangle.
This is a saddle point for the free energy as a function of the coordinates $%
(x,y)$ in the plane of the triangle, and the state is unstable. The second
minimum is reached when all the vortices are in the center of the triangle,
and the vortex-antivortex combination degenerates to a giant vortex $L_{2v}=2
$. This local minimum represents a metastable state.

\begin{figure}
\protect\centerline{\epsfbox{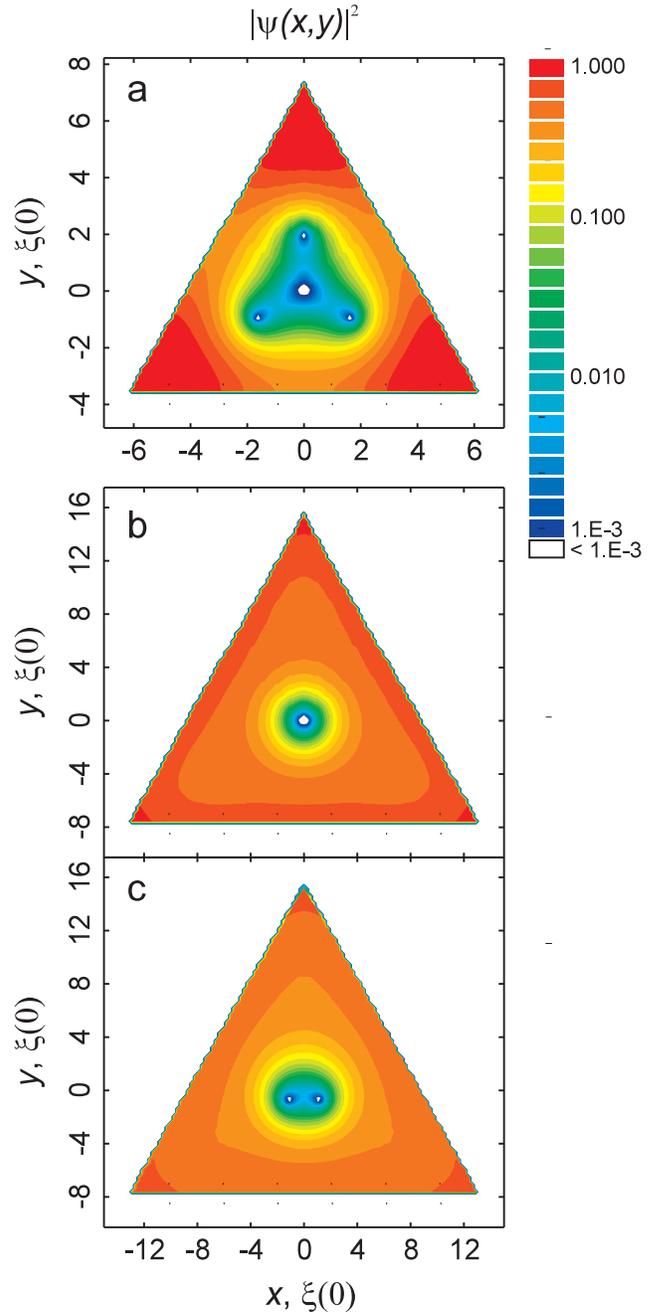}}
\smallskip 
\caption{
The distribution of the squared modulus of the order
parameter $\mid\psi(x,y)\mid^{2}$ for the states with the total vorticity $%
L=2$ in mesoscopic superconducting triangles at $T/T_{c}=0.96$, $%
H_{0}=0.32H_{c}(0)$: $3v+1av$ molecule in a type I superconducting triangle
with $\kappa=0.48$ (a); a giant vortex state in a type II superconducting
triangle $\kappa=1.28$ (b); a stable multivortex state in a type II
superconducting triangle (c).
}
\label{Fig.3}
\end{figure}

The absolute minimum is reached when three vortices are situated between the
center and the apexes of the triangle at $1.83\xi(0)$ from the antivortex in
the center (Fig.~2). {\it This vortex-antivortex molecule is
thermodynamically stable.} Its stability can be understood in the following
way. The distribution of the squared modulus of the order parameter $%
\mid\psi(x,y)\mid^{2}$, which relates to the above stable
vortex-antivortex molecule, is shown in Fig.~3a. Four zeros of
$\mid\psi(x,y)\mid^{2}$ correspond to three vortices and one
antivortex. The distribution of the
magnetic field $H(x,y)$ consistent with the above pattern of $%
\mid\psi(x,y)\mid^{2}$ will be presented elswhere\cite{wetrun}. The function
$\mid\psi(x,y)\mid^{2}$ reaches its maximum value in the corners. These
``islands'' of the superconducting phase in the corners prohibit vortices,
which are repelled by an antivortex in the center, from leaving the triangle
through the corners. Thus, vortices, being confined in a type I
superconducting triangle and interacting with an antivortex, form a stable
vortex-antivortex molecule.

It is worth noting, that for our type I sample a strongly enhanced
nucleation field appears due to the confinement of the superconducting
condensate in the mesoscopic triangle. In fact, this provides the ``soft''
scenario for the nucleation of the order parameter, like in bulk type II
superconductors.

\begin{figure}
\protect\centerline{\epsfbox{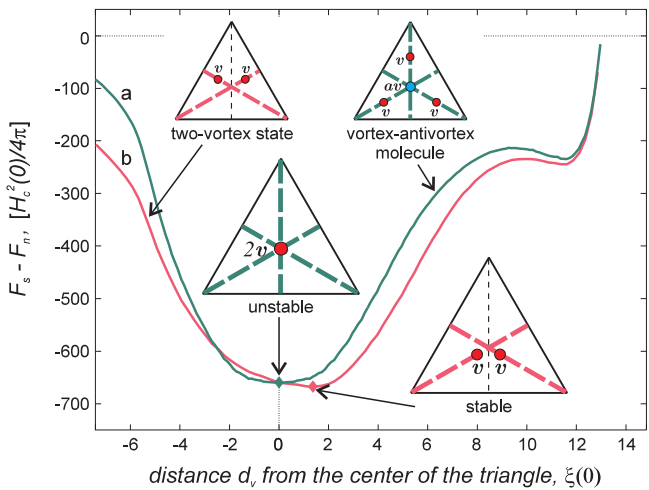}}
\smallskip 
\caption{
The free energy $F_{s}-F_{n}$ [measured in $%
H_{c}^{2}(0)/4\pi$] for the state with the total vorticity $L=2$ as a
function of $d_{v}$ for a mesoscopic type II superconducting triangle at $%
T/T_{c}=0.96$, $H_{0}=0.32H_{c}(0)$, $\kappa=1.28$: $3v+1av$ molecule (a); a
two-vortex state with vortices situated at two different bisectors of the
triangle (b).
}
\label{Fig.4}
\end{figure}

The stable vortex-antivortex patterns are qualitatively different in case of
a type II superconducting triangle. In Fig.~4, the free energy of the ``$3-1$%
'' combination is shown (curve ``a'') as a function of the distance $d_{v}$
from the center of the triangle to vortices. The lowest minimum is reached
when all the vortices are in the center of the triangle. This means that a
giant vortex with vorticity $L_{2v}=2$ is energetically more favorable in a
type II superconducting triangle. (The local minimum at $d_{v} = 11.57\xi(0)$
represents an unstable state.) The corresponding distribution of the squared
modulus of the order parameter $\mid\psi(x,y)\mid^{2}$ is plotted in
Fig.~3b. Although a giant vortex state with $L_{2v}=2$ has a lower free
energy than the ``$3-1$'' combination, the equilibrium is reached for a
multivortex state when two vortices are at two different bisectors (cf. Ref.~%
\onlinecite{bp}) of the triangle (see Fig.~4, curve ``b''). The order
parameter pattern corresponding to this stable two-vortex state is shown in
Fig.~3c.

In conclusion, we have found deep in the superconducting state a
thermodynamically {\it stable} vortex-antivortex configuration for a
mesoscopic type I superconducting triangle, although until now it has been
thought that vortex-antivortex patterns are unstable and they can manifest
themselves only in the close vicinity to the phase boundary.
Vortex-antivortex arrays become unstable in a type II superconducting
triangle, in accordance with previous reports. The stability of the
vortex-antivortex molecules in type I superconducting triangles is due to
the change of the sign in the vortex-vortex and vortex-antivortex
interaction forces when passing through the dual point $\kappa=1/\sqrt{2}$,
combined with the condensate confinement by the boundaries of the mesoscopic
triangle.

This work has been supported by GOA BOF UA 2000, IUAP, the FWO-V projects
Nos. G.0306.00, G.0274.01, WOG WO.025.99N (Belgium), and the ESF Programme
VORTEX. Useful discussions with L.~Van~Look, M.~Morelle and G.~Teniers are
acknowledged.

\end{document}